\documentclass[preprint,11pt]{elsarticle}

\usepackage{amssymb}
\usepackage{amsmath}
\journal{Optics and Lasers in Engineering}

\begin{document}

\begin{frontmatter}

%% Title, authors and addresses

%% use the tnoteref command within \title for footnotes;
%% use the tnotetext command for theassociated footnote;
%% use the fnref command within \author or \affiliation for footnotes;
%% use the fntext command for theassociated footnote;
%% use the corref command within \author for corresponding author footnotes;
%% use the cortext command for theassociated footnote;
%% use the ead command for the email address,
%% and the form \ead[url] for the home page:
%% \title{Title\tnoteref{label1}}
%% \tnotetext[label1]{}
%% \author{Name\corref{cor1}\fnref{label2}}
%% \ead{email address}
%% \ead[url]{home page}
%% \fntext[label2]{}
%% \cortext[cor1]{}
%% \affiliation{organization={},
%%             addressline={},
%%             city={},
%%             postcode={},
%%             state={},
%%             country={}}
%% \fntext[label3]{}

\title{Water immersion single-mirror schlieren imaging system for flow visualization}

%% use optional labels to link authors explicitly to addresses:
%% \author[label1,label2]{}
%% \affiliation[label1]{organization={},
%%             addressline={},
%%             city={},
%%             postcode={},
%%             state={},
%%             country={}}
%%
%% \affiliation[label2]{organization={},
%%             addressline={},
%%             city={},
%%             postcode={},
%%             state={},
%%             country={}}

\author[1]{Shubham Saxena\fnref{label1}} %% Author name
%% Author affiliation
\affiliation[1]{organization={Department of Physics, Indian Institute of Technology Delhi},%Department and Organization
            addressline={Hauz Khas}, 
            city={New Delhi},
            citysep={}, % Uncomment if no comma needed between city and postcode
            postcode={110016}, 
            % state={},
            country={India}}
\fntext[label1]{Currently with Institute of Photonics and Quantum Sciences, Heriot-Watt University, Edinburgh EH14 4AS, United Kingdom.}
            
\author[2]{Manish Kumar\corref{cor1}} %% Author name
\ead{kmanish@iitd.ac.in}
\cortext[cor1]{Corresponding author.}
%% Author affiliation
\affiliation[2]{organization={Centre for Sensors, Instrumentation and Cyber-Physical System Engineering (SeNSE), Indian Institute of Technology Delhi},%Department and Organization
            addressline={Hauz Khas}, 
            city={New Delhi},
            citysep={}, % Uncomment if no comma needed between city and postcode
            postcode={110016}, 
            % state={},
            country={India}}

%% Abstract
\begin{abstract}
%% Text of abstract
Schlieren imaging is a popular optical technique for visualizing flow in transparent media. In-water high-sensitivity flow visualization, using schlieren imaging, is usually performed with a large-footprint two-mirror z-configuration. Here, we present a small footprint, easy-to-implement, single-mirror schlieren imaging system for in-water flow visualization. The same system is capable of high-sensitivity flow visualization in air as well. At its core, our system uses a concave mirror with water immersion. We present theoretical analysis and experimental results to show that this water immersion helps reduce the system's footprint by 25\%. Our water immersion-based single-mirror schlieren imaging method additionally reduces mirror surface artifacts, increasing the sensitivity of flow visualization. This technique enables a low-cost schlieren system, as demonstrated experimentally using an inexpensive concave mirror. We also provide the experimental validation of high sensitivity in-water flow visualization for some transparent chemicals or solutions.
\end{abstract}

%%Graphical abstract
% \begin{graphicalabstract}
%\includegraphics{grabs}
% \end{graphicalabstract}

%%Research highlights
% \begin{highlights}
% \item Immersing the mirror in water minimizes the size of the single-mirror schlieren imaging system.
% \item Water immersion decreases the schlieren artifacts created by mirror surface irregularities.
% \item With water immersion, a cheap concave mirror can be used for sensitive flow visualization.
% \item The water immersion single-mirror schlieren imaging technique is highly sensitive to both in-air and in-water flow visualization.
% \end{highlights}

%% Keywords
\begin{keyword}
%% keywords here, in the form: keyword \sep keyword
schlieren imaging\sep flow visualization\sep water immersion
%% PACS codes here, in the form: \PACS code \sep code

%% MSC codes here, in the form: \MSC code \sep code
%% or \MSC[2008] code \sep code (2000 is the default)

\end{keyword}

\end{frontmatter}

%% Add \usepackage{lineno} before \begin{document} and uncomment 
%% following line to enable line numbers
%% \linenumbers

%% main text
%%

%% Use \section commands to start a section
\section{Introduction}
\label{intro}
%% Labels are used to cross-reference an item using \ref command.

%%%%%%%%%%%%%%%%%%%%%
Schlieren imaging visualizes refractive index gradients in a transparent media by converting small angular deflections of light beam into corresponding intensity variations using a cutoff element \cite{settles2001schlieren,mercer_optical_2003}.
In spite of its simple design, it offers high sensitivity to refractive index variations. Schlieren imaging has been widely applied to investigate convection and mixing phenomena in transparent media \cite{barnes1945schlieren, alvarez2009temperature, tanda2014heat}. The applicability of schlieren to in-liquid flows visualization was demonstrated early on, including water tunnel experiments visualizing refractive gradients due to temperature differences \cite{fiedler1985schlieren}. In recent years, schlieren has been explored in multiple liquid media based studies such as microfluidic mixing \cite{sun2013quantitative}, natural and thermally driven convection \cite{okhotsimskii1998schlieren, babich2023situ}, and dissolution-induced concentration fields in multiphase environments \cite{amarasinghe2021co2}. It has also been employed for quantitative studies, including heat-transfer measurements in water \cite{tanda2014heat}, temperature-field reconstruction in buoyancy-driven liquid convection \cite{srivastava2004imaging}, and concentration-gradient analysis in liquid convection and crystal growth systems \cite{gupta2010color}. In such liquid media based flow visualization studies, schlieren provides non-intrusive observation of evolving refractive index distributions that cannot be readily captured using contact-based probes without disturbing the flow \cite{tanda2014heat}.

\begin{figure}[t]
\centering\includegraphics[width=12.5cm]{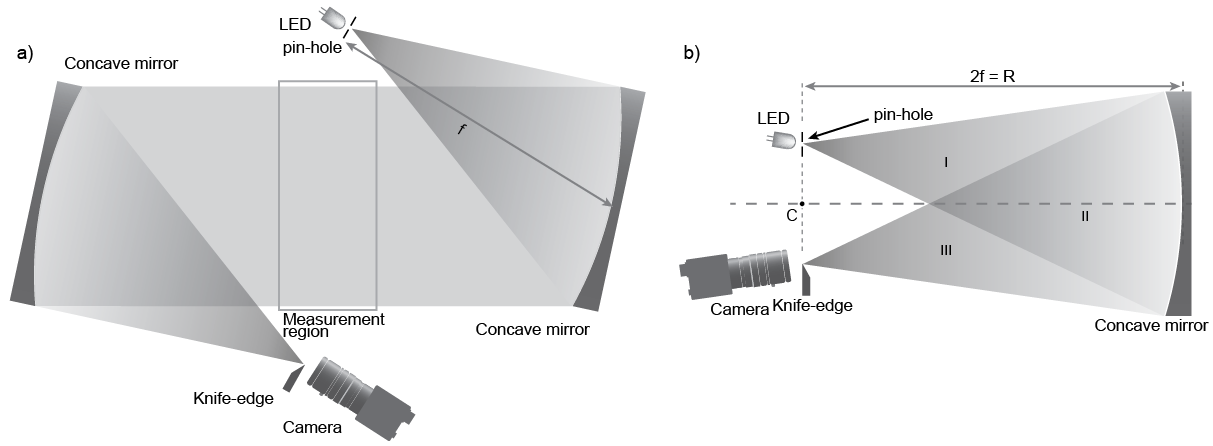}
\caption{\label{fig:1}Comparison between two schlieren configurations: a) two-mirrors based z-configuration and b) single-mirror configuration.}
\end{figure}

In practice, mirror-based schlieren systems for liquid flows are most commonly implemented using the classical two-mirror z-type configuration \cite{settles2001schlieren, mercer_optical_2003, tanda2014heat}. This arrangement employs two concave mirrors separated by several focal lengths to generate and refocus a collimated beam (see Fig. \ref{fig:1}a), providing strong contrast control and a well-defined measurement region \cite{settles2001schlieren, hargather2012comparison, maharjan2023design}. The z-type schlieren system can also use off-axis parabolic mirror pairs to avoid astigmatism \cite{zheng2022methodology}. However, the use of two precision mirrors and a long optical path leads to a physically large setup which requires careful alignment \cite{mercer_optical_2003}. This limits portability and increases both system complexity and cost. These conventional schlieren systems often require large optical tables and extended light paths, which limits their practicality for compact laboratory environments and small-scale liquid flow visualization experiments \cite{tong2026compact}. The high sensitivity of schlieren also demands for high precision mirror surfaces finish. Because schlieren contrast is proportional to angular deflection at the cutoff plane, this elevated sensitivity renders even minor mirror imperfections directly visible. These defects, such as dents or surface irregularities, interfere with the visualization of the flow structures, and obscure the image features of interest. This is why the realization of high sensitivity schlieren imaging relies on the use of high precision first surface telescope mirrors \cite{settles2001schlieren}. 

Background-oriented schlieren (BOS) and related synthetic schlieren approaches provide a simpler alternative for flow visualization by avoiding precision mirrors and instead tracking the apparent displacement of a background pattern using digital image correlation \cite{raffel2015background, moisy2009synthetic, richard2001principle, rabha2025pocket}. However, because BOS relies on a fixed window size-based numerical computation, its capacity to resolve weaker density gradients that lead to smaller displacements than the window size is limited \cite{settles2001schlieren,schmidt2025twenty}. As a result, it often underperforms traditional mirror-based schlieren when targeting weak refractive-index gradients in liquids \cite{schmidt2025twenty,fisher2019experimental}.

Single-mirror schlieren configuration provides both a compact setup and high-resolution flow visualization. It eliminates one of the mirrors required in the z-type setups and places the source and cutoff elements at twice the focal length of a concave mirror (see Fig. \ref{fig:1}b) \cite{taylor1933improvements,settles2018smartphone}.  In this geometry, the light beam passes through the test region twice, effectively doubling the angular deflection and increasing sensitivity \cite{pastor2007evaporating,gena2020qualitative}. Such setups are used in laboratory-scale liquid flow studies where reduced system size and simplified alignment are desirable \cite{ settles2018smartphone}. This makes single-mirror configurations particularly appealing for small laboratory installations. This single-mirror schlieren imaging arrangement should be adequate for in-water flow visualization. However, researchers have been focused on using either a z-type or a BOS approach for in-water flow visualizations. The primary explanation for this could be that a single-mirror schlieren imaging setup uses converging and diverging rays in the test region \cite{barnes1945schlieren}. This indicates that the scale or magnification factor of any flow varies with its distance from the mirror. If a schlieren setup is totally submerged in water, it will produce an indecipherable flow image due to the overlap of many scaled flow patterns along the journey. This issue could be resolved for a flow contained in a thin region at a fixed distance from the mirror. This forms the basis of the work we present here. 

We propose water immersion of the concave mirror and formation of a vertical single-mirror schlieren configuration for in-water flow visualization. Water or liquid immersion is well known in the context of commercial microscope objectives where a high index liquid immersion design improves the numerical aperture. This approach is also widely used for designing custom microscopy systems \cite{voigt2024reflective}. Liquid immersion is also helpful in using 3D printed diffractive optical elements for visible wavelength based applications \cite{orange20213d}. In the context of a single-mirror schlieren imaging, the water forms a thin (or sometimes a thick) water lens, which stays in contact with the mirror. Immersing the concave mirror in water increases its effective optical power through refraction at the water–air interface, reducing the effective focal length and shortening the physical system length. At the same time, the increased surrounding refractive index reduces the apparent contrast of surface-induced artifacts, improving image quality even when using mirrors with modest surface accuracy. This enables a compact and low-cost implementation without compromising the image quality. We present a theoretical analysis of the immersion-induced modification of optical power and experimentally demonstrate a reduction in system footprint together with improved robustness to mirror surface defects. Since the measurement region is very close to the mirror surface, this approach maximizes the field-of-view attainable using single-mirror schlieren imaging. To avoid any potential damage to the delicate front surface mirror during immersion, we also show how an inexpensive concave mirror can be used for liquid immersion schlieren imaging. This method also highlights the accessibility of the proposed approach for classroom demonstration as well as rugged industrial applications.

\section{Theory}
\label{theory}
In this work, we focus on a single-mirror schlieren imaging system for in-water flow visualization. Immersing a concave mirror in liquid (for example, water) has two main advantages. First, it causes a further reduction in the footprint of the schlieren system, and second, it leads to a reduction in the imaging artifacts induced by mirror surface irregularities or defects. In the following, we elaborate on these two effects in the schlieren imaging system with a proper theoretical explanation.

\subsection{Schlieren system footprint reduction}
\label{footprint-theory}
Two-mirror based z-configuration is the most popular schlieren imaging configuration. Figure~\ref{fig:1}a shows the schematics for this z-configuration.  Here, a LED or any other light source is placed next to a pin hole to create tiny point illumination.  This light illuminates a concave mirror kept focal length apart in an off-axis configuration. The light rays get collimated upon mirror reflection. These collimated rays are then reflected off a similar mirror to converge back to a point. A knife-edge is placed here as a schlieren cut-off element. A camera system beyond the knife-edge is then responsible for imaging. Here, the region between two mirrors is the measurement region for flow visualization. This two-mirror z-configuration is bulky, difficult to align, and expensive to set-up.

There exists a single mirror schlieren imaging configuration that mitigates most of the drawbacks of z-configuration. This compact configuration is shown in Fig.~\ref{fig:1}b. Here, the light-source and cut-off elements are both twice the focal length apart from the mirror. This configuration is naturally much more compact in comparison to the standard z-configuration. Since only one mirror is involved, the optical alignment is straight forward. However, this configuration does have a few drawbacks. The off-axis arrangement of the light source and cut-off element entails that the measurement region is divided into three parts. Region II is a natural choice for measurement region, but this is a double-pass region which may lead to the appearance of double schlieren images. On the other hand, region I and III are identical where one may observe single-view schlieren imaging. However, these regions are very small and not appropriate for a large field of view schlieren imaging. The water immersion of a concave mirror enables flow visualization at a very small distance from the mirror. In this situation, the region II becomes double image free.   

\begin{figure}[t]
\centering\includegraphics[width=10cm]{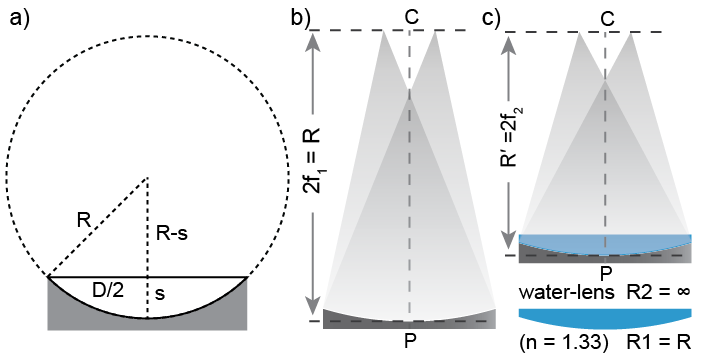}
\caption{\label{fig:2} Effect of a thin liquid lens on the radius of curvature of a concave lens. a) Sag of a concave mirror, b) radius of curvature of a concave mirror, and c) effective radius of curvature of a concave mirror filled with water. The water-lens is a plano-convex lens with the convex side having same but opposite radius of curvature as the mirror.}
\end{figure}

Clearly, the single-mirror configuration offers a compact setup for performing schlieren imaging. Here, the system length is twice the mirror focal length. Our idea of using liquid immersion can make the single mirror schlieren imaging system more compact. For a thorough analysis of this concept we first consider a spherical concave mirror as shown in Fig.~\ref{fig:2}a. Let $R$ be the radius of curvature and let $D$ be the aperture size. Let $s$ represent the sag in the mirror. Then this sag $s$ can be expressed as \cite{hecht2012optics}:

\begin{equation}
 s = R-\sqrt{R^2-\left(\frac{D}{2}\right)^2} = R-R\left[{1-\left(\frac{D}{2R}\right)^2}\right]^{1/2}.
 \label{eq:1}
\end{equation}

Since $\left(\frac{D}{2R}\right) < 1$ and $\left(\frac{D}{2R}\right)^2 << 1$, we can use the \textit{binomial expansion} of Eq.~(\ref{eq:1}) and truncate the higher order terms to obtain:
\begin{equation}
 s \approx R-R\left({1-\frac{D^2}{8R^2}}\right) = \frac{D^2}{8R} = \frac{D^2}{16f} = \frac{D}{16\times F/\#}.
 \label{eq:2}
\end{equation}

The $F/\#$ of mirrors used in schlieren imaging is typically greater than four. Thus, sag $s$ is smaller than $D/64$. We have calculated the values of sags for the mirrors used in this work, and have listed them in Table~\ref{tab:table1}. Clearly, the value of sag is much smaller than mirror aperture or radius of curvature. This implies that when some liquid is poured to fill the mirror, it forms a thin liquid lens whose radius of curvatures is $R1 = R$ and $R2 = \infty$, where $R$ is the radius of the curvature of the mirror. 

% \smallskip
\begin{table}
\caption{\label{tab:table1}Sag of the mirrors used in this work}
\centering
\begin{tabular}{ccccc}
\hline
S.No. & R (in mm) & D (in mm) & F/\# & Sag s (in mm)\\
\hline
i & 800 & 75 & 5.33 & 0.9\\
ii & 400 & 50 & 4.0 & 0.8\\
\hline
\end{tabular}
\end{table}

The presence of this thin liquid lens reduces the effective radius of curvature of the concave mirror. Figure~\ref{fig:2} shows both scenarios side by side: a bare spherical mirror (in Fig. \ref{fig:2}b) and a mirror with liquid lens ((in Fig. \ref{fig:2}c). For the bare mirror $2f_1 = R$, where $f_1$ is the focal length of the mirror. The power of this mirror is given by:
\begin{equation}
P_1 = \frac{1}{f_1} = \frac{2}{R}.
 \label{eq:3}
\end{equation}

Using \textit{lens maker's} formula for the liquid lens we get:
\begin{equation}
\frac{1}{f_2} = (n-1) \left[ \frac{1}{R1}-\frac{1}{R2}\right] = (n-1) \left[ \frac{1}{R}-\frac{1}{\infty}\right] = \frac{n-1}{R},
 \label{eq:4}
\end{equation}
where, $f_2$ is the focal length of the liquid lens and $n$ is the refractive index of the given liquid \textit{i.e.}\ water. 
Using Eq.~(\ref{eq:4}), the power of this water-lens can be written as:
\begin{equation}
P_2 = \frac{1}{f_2} = \frac{n-1}{R}.
 \label{eq:5}
\end{equation}

Since the mirror and the liquid lens are in contact, their combined optical power will be an arithmetic sum of the individual elements. We note that the liquid lens power gets added twice as a result of the mirror reflection. Thus, the combined power using Eq.~(\ref{eq:3}) and Eq.~(\ref{eq:5}) is:
\begin{equation}
P= P_1 + 2P_2 = \frac{2}{R} + 2 \left(\frac{n-1}{R}\right) = \frac{2n}{R}.
 \label{eq:6}
\end{equation}

Except for an additional multiplicative factor $n$, Eq.~(\ref{eq:6}) is identical to Eq.~(\ref{eq:3}). Therefore, the effective focal length of a mirror with liquid immersion is given by:
\begin{equation}
f_2 = \frac{R}{2n}, 
 \label{eq:7}
\end{equation}
where, $R$ is the radius of curvature of the mirror and $n$ is the refractive index of immersion media. Since $n > 1$, we have $f_2 < f_1$. The effective radius of curvature of the system with liquid immersion decreases by a factor $n$ and the length of the resultant system is: 
\begin{equation}
R' = 2f_2 = R/n.
 \label{eq:8}
\end{equation}
For water, $n = 1.33$ which leads to $R' = R/1.33 \approx0.75R$ corresponding to a $25\%$ reduction in the effective radius of curvature (and effective focal length).

\begin{figure}[t]
\centering\includegraphics[width=10cm]{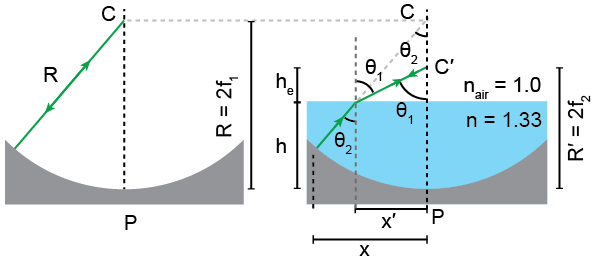}
\caption{\label{fig:3} Effect of a thick liquid lens on the radius of curvature of a concave lens. Concave mirror without water lens (left) and concave mirror with a thick water lens (right).}
\end{figure}

The height of liquid immersion media is not limited to sag of the mirror, it can be much larger when the mirror is completely immersed. Figure~\ref{fig:3}a shows this case of complete immersion compared to a bare mirror. We analyze this case in detail to arrive at the resultant radius of curvature $R'$. As shown in Fig.~\ref{fig:3}a, the center of curvature $C$ shifts to $C'$ due to the refraction at the liquid-air interface. The distance of $C'$ from the pole \textit{i.e.}\ geometric center of the mirror surface, is $R' = 2f_2$. The height of the liquid column is $h$. The effective radius of curvature is a sum of the height of the liquid column and the air gap: $R'=h+h_e$.
 
We consider an optical ray that starts at the effective center of curvature $C'$, gets refracted at the air-liquid interface, and meets the mirror surface $x$ distance apart from its pole. This refracted ray, when traced backwards, virtually meets the principal axis at the center of curvature $C$. Let $\theta_1$ be the angle of incidence of this ray in air and $\theta_2$ be the angle of refraction in liquid. Then, using \textit{Snell's law} we have:
\begin{equation}
\sin\theta_1 = n\times \sin\theta_2 = \frac{nx}{R},
 \label{eq:9}
\end{equation}
where $n$ is the liquid refractive index, and we use basic trigonometry to express $\sin\theta_2$ in terms of offset distance $x$ and radius of curvature $R$. Next, we have two similar triangles with a common vertex $C$ and two side lengths $(x,R)$ \& $(x',R-h)$, respectively. Therefore, we can express the triangle sides ratio as follows:
\begin{eqnarray*}
\frac{x'}{R-h} = \frac{x}{R},
\end{eqnarray*} which can be adjusted to arrive at the following:
\begin{equation}
x'= \frac{x}{R}\times(R-h) = x \left(1-\frac{h}{R}\right).
 \label{eq:10}
\end{equation}
 
Next, we consider the small triangle made of the optical ray that intersects the vertex point $C'$, the principal axis of the mirror, and the liquid surface. In this triangle, we have $\tan\theta_1 = x'/h_e$ or, $h_e = x'/\tan\theta_1 = x'\cos\theta_1/\sin\theta_1$. Using Eq.~(\ref{eq:9}) and Eq.~(\ref{eq:10}), this expression can be expanded as:
\begin{equation}
h_e = x \left(1-\frac{h}{R}\right)\times \frac{\sqrt{1-\frac{n^2x^2}{R^2}}}{\frac{nx}{R}} = \frac{R}{n}\left(1-\frac{h}{R}\right) \left(1-\frac{n^2x^2}{R^2}\right)^{1/2}.
 \label{eq:11}
\end{equation}
The last term of Eq.~(\ref{eq:11}) can be simplified by using \textit{binomial expansion} and truncating higher-order terms. This is possible because $x \leq D/2$ and $D<R$, which implies $nx/R < 1$ and $n^2x^2/R^2 << 1$. So, the last term in Eq.~(\ref{eq:11}) becomes $[1-n^2x^2/(2R^2)] = [1-n^2x^2/(2R^2)]$. This term depends on the value of $x$. For the minimum value $x=0$, the term is reduced to unity. For, the maximum value $x=D/2$, the term becomes $[1-n^2D^2/(8R^2)] = [1-n^2D^2/(32f^2)] = [1-n^2/(32(F/\#)^2)]$. For our schlieren imaging setup, since $F/\# \geq 4$ and $n = 1.33$, we get $n^2/(32(F/\#)^2) \leq (1.33)^2/(32\times 16) = 0.003$. Clearly, this value is so small that $[1-n^2D^2/(8R^2)] \approx 1$. Thus, Eq.~(\ref{eq:11}) reduces to:
\begin{equation}
h_e = \frac{R}{n}\left(1-\frac{h}{R}\right).
 \label{eq:12}
\end{equation}
Finally, the radius of curvature becomes: 
\begin{equation}
R' = h_e + h = \frac{R}{n}\left(1-\frac{h}{R}\right) + h = \frac{R}{n} + h\left(1-\frac{1}{n}\right).
 \label{eq:13}
\end{equation}
For no liquid immersion \textit{i.e.} $n = 1$, this equation reduces to $R' = R$ as expected. We can find the extreme values for $R'$ by considering two special cases. First, when $h \to 0$, we get $R' = R/n$. This is exactly same as Eq.~(\ref{eq:8}) corresponding to the case of thin liquid lens. Second, when $h = R$, we get $R' = R$ \textit{i.e.}\ no reduction in the effective radius of curvature. This is expected as the liquid column occupying the whole height implies that light rays do not encounter any liquid-air interface to bend at. This does not result in a change in the effective radius of curvature of the system. This consideration also proves that the effective radius of curvature for a thick liquid lens-based concave mirror is smaller than the original radius of curvature of the mirror. 

\subsection{\label{sec:level2B}Mirror surface artifacts reduction}
\label{artifacts-theory}

\begin{figure}[t]
\centering\includegraphics[width=10cm]{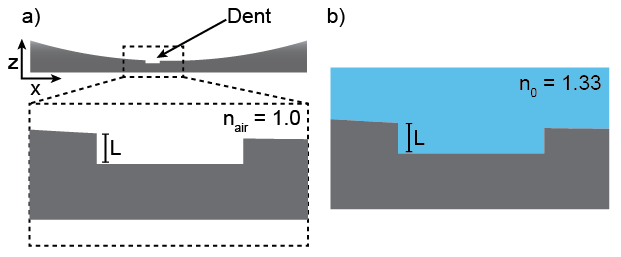}
\caption{\label{fig:4} Effect of liquid immersion on surface artifact reduction. a) A simplified dent in a mirror surface having depth $L$, and b) liquid (water) immersion of the dented mirror.}
\end{figure}

Mirror surface artifacts \textit{e.g.} dents, bumps, or scratches are highly visible in schlieren imaging and pose challenges in obtaining a clean flow visualization of the fluid of interest. The origin of these high-contrast artifacts lies in the basic principles of schlieren imaging. Schlieren imaging contrast is directly dependent on the amount of light cut-off by the knife edge, which in turn depends on the bending of light rays. The ray curvature due to an optical inhomogeneity is given by the following expressions \cite{settles2001schlieren}:
\begin{subequations}
\label{eq:14}
\begin{equation}
\frac{\partial^2 x}{\partial z^2} = \frac{1}{n}\frac{\partial n}{\partial x}, 
\end{equation}
\begin{equation}
\frac{\partial^2 y}{\partial z^2} = \frac{1}{n}\frac{\partial n}{\partial y}, 
\end{equation}
\end{subequations}
where $z$ is the optical propagation direction of the undisturbed rays, $x$ and $y$ are orthogonal directions in the Cartesian coordinates along which we have some optical inhomogeneity, and $n$ is the local refractive index of the given schlieren sample or media. Integrating Eq.~(\ref{eq:14}) once we obtain optical ray bending angle components $\epsilon_x$ and $\epsilon_y$: 
\begin{subequations}
\label{eq:15}
\begin{equation}
\epsilon_x = \frac{1}{n} \int \frac{\partial n}{\partial x} \partial z = \frac{L}{n_0}\frac{\partial n}{\partial x}, 
\end{equation}
\begin{equation}
\epsilon_y = \frac{1}{n} \int \frac{\partial n}{\partial y} \partial z = \frac{L}{n_0}\frac{\partial n}{\partial y}, 
\end{equation}
\end{subequations}
where $L$ is the extent of the schlieren object along the optical axis \textit{i.e.} $z-axis$, and $n_0$ is the refractive index of the surrounding medium. This expression explains how schlieren visualization arises from the refractive index gradient $\partial n/ \partial x$ and $\partial n/ \partial y$ terms. We use the above equation to analyze the contrast of an artifact that arises in a spherical mirror with a rectangular dent (see Fig.~\ref{fig:4}). We have a large refractive index gradient $\partial n/\partial y$ at the edges of this dent. This makes it highly visible in schlieren images. Moreover, the depth of the dent $L$ is also responsible for increasing its contrast in the schlieren image, making it difficult to measure any fluid flow that overlaps the dent. 

While Eq.~(\ref{eq:15}) describes the contrast observed due to a dent in terms of refractive index gradients $\partial n/\partial x$ \& $\partial n/\partial y$ and depth $L$, it also contains a less considered term $n_0$ \textit{i.e.} refractive index of the surrounding medium. A liquid immersion of this dent-containing mirror increases the value of $n_0$, thus reducing the angle of bend, which leads to a lower schlieren contrast of the dent. For water immersion $n_0 = 1.33$, this implies $\approx 25\%$ reduction in contrast to the dent artifact. In addition to dents, the same treatment is valid even for artifacts induced by bumps, irregular glass deposits, or the uneven surface of the base or walls of any sample containers.

\section{Experimental results}
\label{exp}

We now go through various experimental demonstrations for the theory presented above. We start by showing that the water immersion reduces the effective setup size for a single mirror schlieren imaging system. Next, we demonstrate that water immersion reduces the mirror surface artifacts. Finally, we show that water immersion single mirror schlieren system provides a high sensitivity platform for flow visualization in water.

\subsection{\label{sec:level3A}System footprint reduction}

\begin{figure}[t]
\centering\includegraphics[width=13cm]{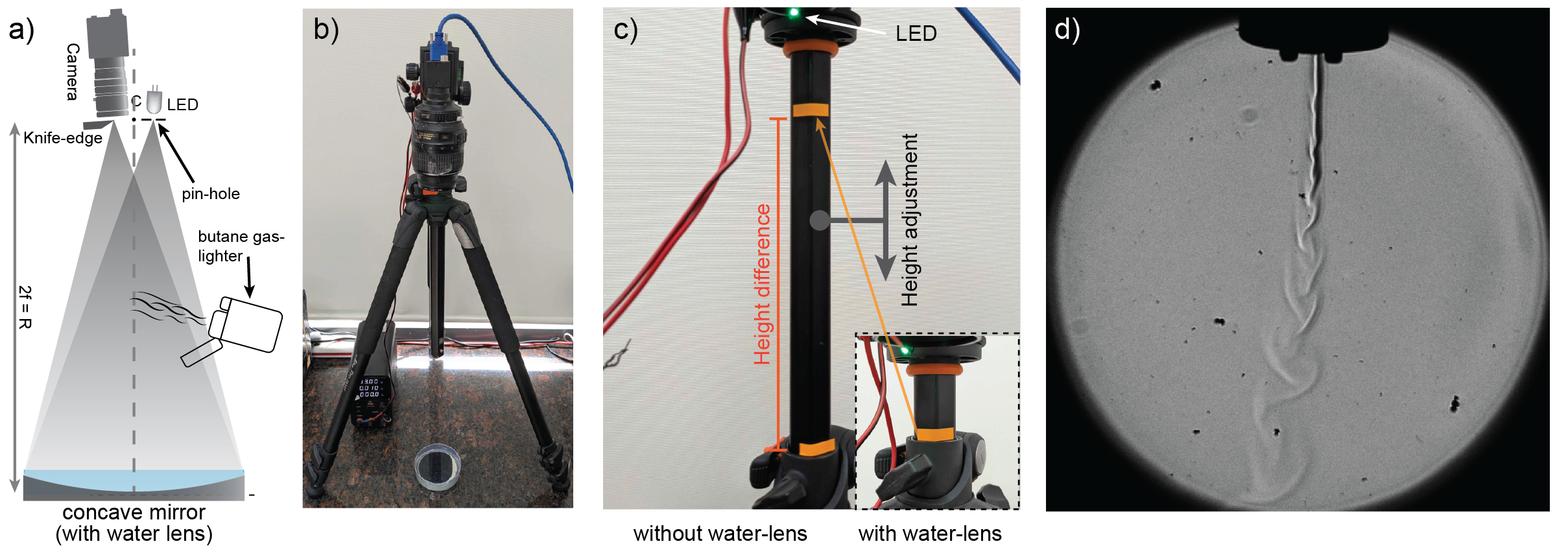}
\caption{\label{fig:5} System footprint reduction with water-lens assisted schlieren. a) Schematics of the single mirror water-lens assisted schlieren setup. b) Actual setup in the lab. c) The tripod height adjustments for schlieren imaging before (main image) and after (inset image) water-immersion of the lens. d) Confirmation of the schlieren contrast with visualization of the butane gas ejecting from a lighter.}
\end{figure}

We built a schlieren imaging setup in a vertical geometry as per the schematics shown in Fig. \ref{fig:5}a. We used a 75 mm diameter telescope mirror having 400 mm focal length, a tiny surface mount style LED, a machine vision camera (BFS-U3-16S2C-CS, Teledyne FLIR), a zoom camera lens (18-55 mm AF-S DX, Nikon), and mounted them together on a tripod (Alta Pro 263AT, Vanguard). The LED was attached at the front surface of the camera lens without blocking its field of view. The finished setup is shown in Fig. \ref{fig:5}b. We found it more convenient to use the camera lens-iris, instead of an external knife-edge, as the cut-off element \cite{kannan2020schlieren}. We used butane gas from a butane-lighter as the test subject. Next, we aligned the schlieren system to obtain schlieren contrast in two configurations: the first being a traditional approach without water immersion of the mirror, and the second approach with water immersion of the mirror. This second approach required significant lowering of the camera-LED assembly using the tripod height adjustment rod (shown in Fig. \ref{fig:5}c). Both optimal heights were determined based on the realization of schlieren contrast for the butane gas. Figure \ref{fig:5}d shows the schlieren contrast obtained with water immersion configuration.  The height difference we obtained in our case was $\sim20$ cm. Thus, the resultant radius of curvature $R'$ is $800-200=600$ mm. This is a $25\%$ reduction in the radius of curvature which is consistent with Eq.~(\ref{eq:8}) proving the agreement between theory and experiment. Hence, water immersion is advantageous due to the schlieren system footprint reduction. 

\subsection{\label{sec:level3B}Mirror surface artifacts reduction}

\begin{figure}[t]
\centering\includegraphics[width=12.5cm]{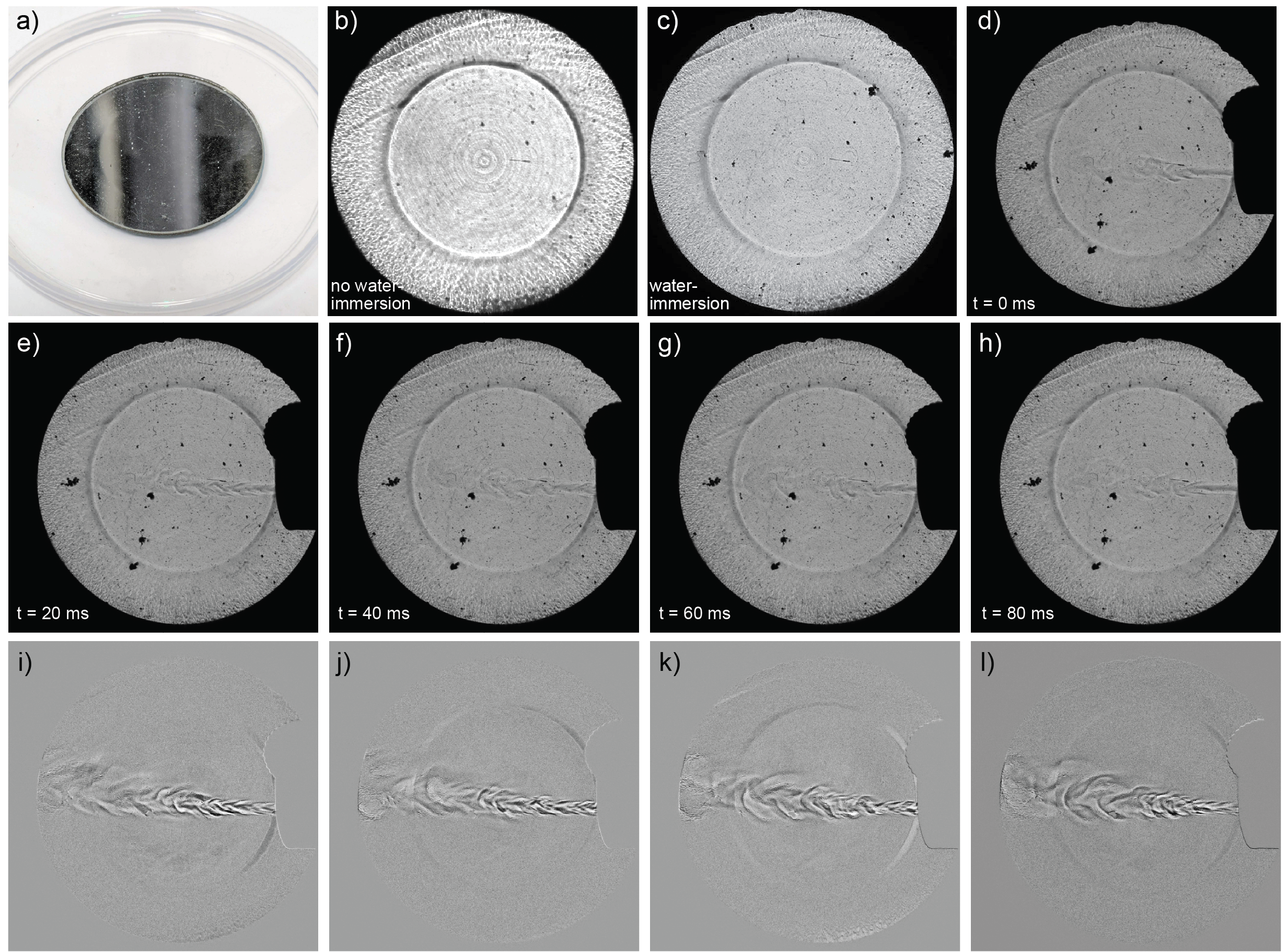}
\caption{\label{fig:6} Schlieren imaging with a low-cost concave mirror. a) A low-cost concave mirror. b) Schlieren imaging highlights the surface irregularities in this mirror. c) Water immersion with the same mirror suppresses surface artifacts. d-h) Five sequential frames showing schlieren imaging of butane gas flow over the water immersion single mirror schlieren setup. i-l) Consecutive frame subtraction and contrast enhancement of the five frames shown in d-h.}
\end{figure}

The next experiment is targeted towards demonstrating another advantage of the water immersion \textit{i.e.} mirror surface irregularity induced schlieren artifact contrast reduction. For this demonstration we used a cheap (less than one hundred INR or less than one USD) educational grade concave mirror (diameter 50 mm and focal length 200 mm). Figure \ref{fig:6}a shows the mirror along with the optimally configured schlieren imaging obtained with the dry mirror and no sample (in Fig. \ref{fig:6}b). Clearly, the schlieren image shows every zonal and surface defects in the mirror. Such a mirror is hardly usable in schlieren imaging. Next, we immersed this mirror in water and readjusted the setup to provide schlieren contrast. As seen in Fig. \ref{fig:6}c, the contrast of defect sites reduces significantly to demonstrate the another advantage of water immersion theoretically explained above. To drive the application aspect of this observation, we extended the experiment and recorded butane gas flow in a sequence of five frames with 10 ms exposure time and 20 ms time interval (see Fig. \ref{fig:6}d-h). The setup was identical to the one shown in Fig.\ref{fig:5} with only difference being the use of a inexpensive mirror instead of the precision front-surface telescope mirror. The direct visualization of flow is not clear enough in the above setup. However, the flow visualization can be significantly enhanced by a simple consecutive frame subtraction method \cite{manish_SplSub_2026} followed by contrast stretching. This approach is not usable for a quantitative flow measurement but is valuable for its demonstrated sensitivity for a qualitative flow visualization with such a inexpensive mirror. This has a potential to make schlieren imaging easily accessible in low-resource educational laboratory. 

\subsection{\label{sec:level3C}In-water flow visualization} 

\begin{figure}[t]
\centering\includegraphics[width=10cm]{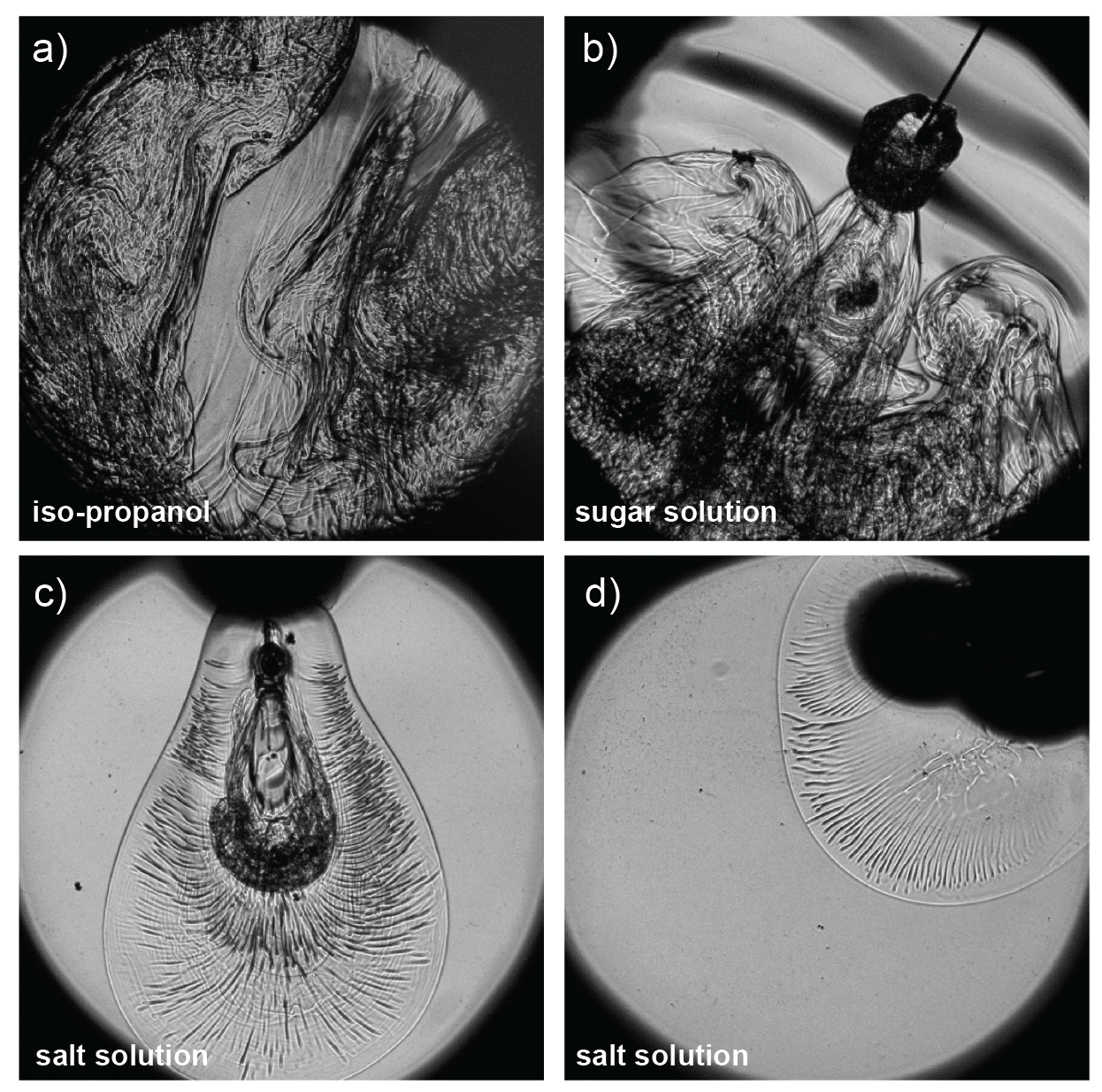}
\caption{\label{fig:7} High sensitivity in-water flow visualization using water immersion single mirror schlieren imaging. a) Isopropanol injected using a syringe. b) Sugar solution injected using a syringe. c) Epsom salt solution injected using a syringe. d) Epsom salt solution dispersing from a wet lens tissue paper.}
\end{figure}
Having seen the advantages of water immersion single mirror schlieren imaging system for footprint reduction and mirror surface artifact suppression, we now demonstrate its in-water high sensitivity flow visualization capability. For this, we use three different kind of liquids or solutions: i) isopropanol (n = 1.377), ii) sugar solution (10 mg/ml, n = 1.336), and iii) Epsom salt solution (10 mg/ml, n = 1.335). The refractive index of isopropanol was noted from its datasheet and sugar and salt solution refractive indices were measured using an Abbe refractometer instrument. These liquids and solutions were filled in a syringe fitted with a fine needle, and injected slowly in the water being held by the concave mirror (75 mm dia, 400 mm focal length) in a water-lens assisted schlieren configuration. These in-water flow patterns became clearly visible in the schlieren setup (Fig. \ref{fig:7}a-\ref{fig:7}c). These needle assisted high speed flow patterns are locally turbulent in nature which creates tiny ripples at the water-air interface.  To assess the sensitivity we next performed flow visualization of the natural diffusion of the Epsom salt solution from a wet tissue paper placed at the edge of mirror containing water. The salt solution slowly diffused in the water creating a flow pattern that was clearly visible in the schlieren setup (see Fig. \ref{fig:7}d). 

\section{Discussion and conclusion}
\label{conclusion}
We have shown that water immersion single-mirror schlieren imaging offers many advantages for flow visualization. Both in-air and in-water flow visualizations are improved. The advantages of space saving compact system, reduction in mirror surface artifact contrast, and the ability to perform high resolution in-water flow visualization make this approach useful. Using water immersion, the single-mirror schlieren system length reduces by $25\%$. We have backed up these advantages both theoretically and experimentally. The ability to use a inexpensive mirror for high sensitivity schlieren imaging is poised to open up new application domains. 

One of the disadvantages of traditional schlieren is that they use first surface mirrors which are very delicate. As a result, schlieren visualization is limited to non-contact approaches mostly in air. The usability of standard non-first surface mirrors, as presented in this work, helps expand schlieren imaging to many new areas. It should be possible to extend schlieren to study many chemical reactions, or observe flow in a transparent corrosive media. 
Although we have used water immersion throughout this work, it is possible to perform liquid immersion with index matching. In such a case, any large deformations in the glass surface can be totally removed. Since a thick liquid lens can also be used, it should be possible to place a sample inside a glass container and then place the container in contact with the liquid contained by the mirror. In this way, the sample flow can be completely isolated from the mirror surface.

While in-contact flow measurement has disadvantages in terms of the risking the mirror surface damage, it also offers clear advantages. There is a complete utilization of the available field of view of the mirror. The field of view is as large as the mirror aperture. The another advantage is the potential doubling of the flow visualization sensitivity due to the double pass. Any density gradients in the double pass leads to a sum of the angular deflections leading to better sensitivity. This is particularly helpful for visualizing very weak density gradients. 

We have used a standard machine vision camera in this work. However, this is not a limiting factor. One may extend it to either a higher-end camera or even a smartphone camera \cite{settles2018smartphone}. A smartphone-enabled system may prove to be particularly suitable for classroom demonstrations of high-sensitivity schlieren imaging. A high-speed camera can be used to expand the application to see ultrasonic wave fields \cite{neumann2006schlieren}, shock waves \cite{hargather2010schlieren}, velocity distribution measurements \cite{fu2001detection}, study of plasma jets \cite{tp2026study}, etc. In conclusion, we strongly believe that the water immersion single-mirror schlieren system will get widely adopted due to the multiple advantages it offers at no additional cost. 

\section*{Acknowledgment} MK acknowledges Indian Institute of Technology Delhi and Anusandhan National Research Foundation (ANRF) for startup grants.

 \bibliographystyle{elsarticle-num} 
 \bibliography{references}

@book{settles2001schlieren,
  title={Schlieren and shadowgraph techniques: visualizing phenomena in transparent media},
  author={Settles, Gary S},
  year={2001},
  publisher={Springer Science \& Business Media}
}

@book{mercer_optical_2003,
	address = {Boston, MA},
	title = {Optical {Metrology} for {Fluids}, {Combustion} and {Solids}},
	copyright = {http://www.springer.com/tdm},
	isbn = {9781441953469 9781475737776},
	url = {http://link.springer.com/10.1007/978-1-4757-3777-6},
	language = {en},
	urldate = {2026-01-21},
	publisher = {Springer US},
	editor = {Mercer, Carolyn R.},
	year = {2003},
	doi = {10.1007/978-1-4757-3777-6},
}

@article{sun2013quantitative,
  title={Quantitative analysis of microfluidic mixing using microscale schlieren technique},
  author={Sun, Chen-li and Hsiao, Tzu-hsun},
  journal={Microfluidics and nanofluidics},
  volume={15},
  number={2},
  pages={253--265},
  year={2013},
  publisher={Springer}
}

@article{okhotsimskii1998schlieren,
  title={Schlieren visualization of natural convection in binary gas--liquid systems},
  author={Okhotsimskii, Andrei and Hozawa, Mitsunori},
  journal={Chemical Engineering Science},
  volume={53},
  number={14},
  pages={2547--2573},
  year={1998},
  publisher={Elsevier}
}

@article{babich2023situ,
  title={In-situ measurements of temperature field and Marangoni convection at hydrogen bubbles using schlieren and PTV techniques},
  author={Babich, Alexander and Bashkatov, Aleksandr and Yang, Xuegeng and Mutschke, Gerd and Eckert, Kerstin},
  journal={International Journal of Heat and Mass Transfer},
  volume={215},
  pages={124466},
  year={2023},
  publisher={Elsevier}
}

@article{alvarez2009temperature,
  title={Temperature measurement of air convection using a schlieren system},
  author={Alvarez-Herrera, C and Moreno-Hern{\'a}ndez, D and Barrientos-Garc{\'\i}a, B and Guerrero-Viramontes, JA},
  journal={Optics \& Laser Technology},
  volume={41},
  number={3},
  pages={233--240},
  year={2009},
  publisher={Elsevier}
}

@article{barnes1945schlieren,
  title={Schlieren and shadowgraph equipment for air flow analysis},
  author={Barnes, Norman F and Bellinger, S Lawrence},
  journal={Journal of the Optical Society of America},
  volume={35},
  number={8},
  pages={497--509},
  year={1945},
  publisher={Optical Society of America}
}

@article{amarasinghe2021co2,
  title={CO2 dissolution and convection in oil at realistic reservoir conditions: A visualization study},
  author={Amarasinghe, Widuramina and Fjelde, Ingebret and Guo, Ying},
  journal={Journal of Natural Gas Science and Engineering},
  volume={95},
  pages={104113},
  year={2021},
  publisher={Elsevier}
}

@article{tanda2014heat,
  title={Heat transfer measurements in water using a schlieren technique},
  author={Tanda, Giovanni and Fossa, Marco and Misale, Mario},
  journal={International Journal of Heat and Mass Transfer},
  volume={71},
  pages={451--458},
  year={2014},
  publisher={Elsevier}
}

@article{hargather2012comparison,
  title={A comparison of three quantitative schlieren techniques},
  author={Hargather, Michael J and Settles, Gary S},
  journal={Optics and Lasers in Engineering},
  volume={50},
  number={1},
  pages={8--17},
  year={2012},
  publisher={Elsevier}
}

@phdthesis{maharjan2023design,
  title={Design and Setup of Z-Type Schlieren Imaging System for Flow Visualization},
  author={Maharjan, Salim},
  year={2023},
  school={IOE Pulchowk Campus}
}

@article{moisy2009synthetic,
  title={A synthetic Schlieren method for the measurement of the topography of a liquid interface},
  author={Moisy, Fr{\'e}d{\'e}ric and Rabaud, Marc and Salsac, K{\'e}vin},
  journal={Experiments in Fluids},
  volume={46},
  number={6},
  pages={1021--1036},
  year={2009},
  publisher={Springer}
}

@article{raffel2015background,
  title={Background-oriented schlieren (BOS) techniques},
  author={Raffel, Markus},
  journal={Experiments in Fluids},
  volume={56},
  number={3},
  pages={60},
  year={2015},
  publisher={Springer}
}

@techreport{pastor2007evaporating,
  title={Evaporating diesel spray visualization using a double-pass shadowgraphy/schlieren imaging},
  author={Pastor, Jos{\'e} V and Garc{\'\i}a, Jos{\'e} M and Pastor, Jos{\'e} M and Zapata, L Daniel},
  year={2007},
  institution={SAE Technical Paper}
}

@article{fisher2019experimental,
  title={An experimental sensitivity comparison of the schlieren and background-oriented schlieren techniques applied to hypersonic flow},
  author={Fisher, TB and Quinn, Mark Kenneth and Smith, KL},
  journal={Measurement Science and Technology},
  volume={30},
  number={6},
  pages={065202},
  year={2019},
  publisher={IOP Publishing}
}

@article{fiedler1985schlieren,
  title={Schlieren photography of water flow},
  author={Fiedler, H and Nottmeyer, K and Wegener, PP and Raghu, S},
  journal={Experiments in fluids},
  volume={3},
  number={3},
  pages={145--151},
  year={1985},
  publisher={Springer}
}

@article{srivastava2004imaging,
  title={Imaging of a convective field in a rectangular cavity using interferometry, schlieren and shadowgraph},
  author={Srivastava, Atul and Phukan, Atanu and Panigrahi, PK and Muralidhar, K},
  journal={Optics and lasers in engineering},
  volume={42},
  number={4},
  pages={469--485},
  year={2004},
  publisher={Elsevier}
}

@article{gupta2010color,
  title={Color schlieren deflectometry for characterization of crystal growth processes: KDP and lysozyme},
  author={Gupta, Anamika Sethia and Panigrahi, PK and Muralidhar, K and Gupta, Rajive},
  journal={Journal of crystal growth},
  volume={312},
  number={6},
  pages={817--830},
  year={2010},
  publisher={Elsevier}
}

@article{schmidt2025twenty,
  title={Twenty-five years of background-oriented schlieren: advances and novel applications},
  author={Schmidt, Bryan E and Bathel, Brett F and Grauer, Samuel J and Hargather, Michael J and Heineck, James T and Raffel, Markus},
  journal={AIAA journal},
  volume={63},
  number={12},
  pages={5028--5058},
  year={2025},
  publisher={American Institute of Aeronautics and Astronautics}
}

@article{richard2001principle,
  title={Principle and applications of the background oriented schlieren (BOS) method},
  author={Richard, Hugues and Raffel, Markus},
  journal={Measurement science and technology},
  volume={12},
  number={9},
  pages={1576},
  year={2001},
  publisher={IOP Publishing}
}

@article{tong2026compact,
  title={A Compact Schlieren Optics Device for Imaging Biological Samples},
  author={Tong, Yimeng and Tang, Jay X},
  journal={Bio-protocol},
  volume={16},
  number={1},
  year={2026}
}

@article{settles2018smartphone,
  title={Smartphone schlieren and shadowgraph imaging},
  author={Settles, Gary S},
  journal={Optics and lasers in engineering},
  volume={104},
  pages={9--21},
  year={2018},
  publisher={Elsevier}
}

@article{rabha2025pocket,
  title={Pocket schlieren: a background-oriented schlieren imaging platform on a smartphone},
  author={Rabha, Diganta and Saini, Dinesh and Kumar, Akshay and Kumar, Vimod and Kumar, Manish},
  journal={Experiments in Fluids},
  volume={66},
  number={8},
  pages={147},
  year={2025},
  publisher={Springer}
}

@article{zheng2022methodology,
  title={Methodology of designing compact schlieren systems using off-axis parabolic mirrors},
  author={Zheng, Lingzhi and Susa, Adam J and Hanson, Ronald K},
  journal={Applied Optics},
  volume={61},
  number={16},
  pages={4857--4864},
  year={2022},
  publisher={Optica Publishing Group}
}

@article{taylor1933improvements,
  title={Improvements in the schlieren method},
  author={Taylor, HG and Waldram, JM},
  journal={Journal of scientific instruments},
  volume={10},
  number={12},
  pages={378--389},
  year={1933}
}

@article{gena2020qualitative,
  title={Qualitative and quantitative schlieren optical measurement of the human thermal plume},
  author={Gena, Amayu W and Voelker, Conrad and Settles, Gary S},
  journal={Indoor air},
  volume={30},
  number={4},
  pages={757--766},
  year={2020},
  publisher={Wiley Online Library}
}

@article{voigt2024reflective,
  title={Reflective multi-immersion microscope objectives inspired by the Schmidt telescope},
  author={Voigt, Fabian F and Reuss, Anna Maria and Naert, Thomas and Hildebrand, Sven and Schaettin, Martina and Hotz, Adriana L and Whitehead, Lachlan and Bahl, Armin and Neuhauss, Stephan CF and Roebroeck, Alard and others},
  journal={Nature Biotechnology},
  volume={42},
  number={1},
  pages={65--71},
  year={2024},
  publisher={Nature Publishing Group US New York}
}

@article{orange20213d,
  title={3D printable diffractive optical elements by liquid immersion},
  author={Orange-Kedem, Reut and Nehme, Elias and Weiss, Lucien E and Ferdman, Boris and Alalouf, Onit and Opatovski, Nadav and Shechtman, Yoav},
  journal={Nature communications},
  volume={12},
  number={1},
  pages={3067},
  year={2021},
  publisher={Nature Publishing Group UK London}
}

@book{hecht2012optics,
  title={Optics},
  author={Hecht, Eugene},
  year={2012},
  publisher={Pearson Education India}
}

@inproceedings{kannan2020schlieren,
  title={Schlieren without Knife-edge},
  author={Kannan, BT and Lingeshwar, R and Gousik, R},
  booktitle={IOP Conference Series: Materials Science and Engineering},
  volume={988},
  issue={1},
  pages={012037},
  year={2020},
  organization={IOP Publishing}
}

@article{neumann2006schlieren,
  title={Schlieren visualization of ultrasonic wave fields with high spatial resolution},
  author={Neumann, Thorsten and Ermert, Helmut},
  journal={Ultrasonics},
  volume={44},
  pages={e1561--e1566},
  year={2006},
  publisher={Elsevier}
}

@article{hargather2010schlieren,
  title={Schlieren imaging of loud sounds and weak shock waves in air near the limit of visibility},
  author={Hargather, Michael John and Settles, Gary S and Madalis, Matthew J},
  journal={Shock Waves},
  volume={20},
  number={1},
  pages={9--17},
  year={2010},
  publisher={Springer}
}

@article{fu2001detection,
  title={Detection of velocity distribution of a flow field using sequences of schlieren images},
  author={Fu, Shan and Wu, Yajue},
  journal={Optical engineering},
  volume={40},
  number={8},
  pages={1661--1666},
  year={2001},
  publisher={Society of Photo-Optical Instrumentation Engineers}
}

@article{tp2026study,
  title={Study of gas flow dynamics of helical plumes in a radiofrequency atmospheric pressure plasma jet},
  author={TP, Radhika and Mallick, Aishik Basu and Kumar, Vimod and Kumar, Manish and Kar, Satyananda},
  journal={Physics of Fluids},
  volume={38},
  number={1},
  year={2026},
  publisher={AIP Publishing}
}

@misc{manish_SplSub_2026,
  author = {Kumar, Manish},
  title = {{Special subtract: v1.0.1}},
  year = {2026},
  publisher = {GitHub},
  journal = {GitHub repository},
  howpublished = {\url{https://github.com/BIOS-lab-IITD/special-subtract}},
  doi = {10.5281/zenodo.18721699}
}

%% else use the following coding to input the bibitems directly in the
%% TeX file.

%% Refer following link for more details about bibliography and citations.
%% https://en.wikibooks.org/wiki/LaTeX/Bibliography_Management

% \begin{thebibliography}{00}

% %% For numbered reference style
% %% \bibitem{label}
% %% Text of bibliographic item

% \bibitem{lamport94}
%   Leslie Lamport,
%   \textit{\LaTeX: a document preparation system},
%   Addison Wesley, Massachusetts,
%   2nd edition,
%   1994.
% \end{thebibliography}

\end{document}